\documentclass[12pt]{article}
\usepackage{geometry}
\usepackage{a4}
\usepackage{graphics}
\usepackage{graphicx,subfigure}
\usepackage{subfigure}
\usepackage{epsf}
\usepackage[T1]{fontenc}
\usepackage[utf8]{inputenc}
\usepackage{amsmath}
\usepackage{amssymb}
\usepackage{cite}
\usepackage{tensor}
\begin{document}
\begin{titlepage}
\title{Massless fermions in null Fermi coordinates and an application to analog gravity in $\text{He}^3$-A}
\author{}
\date{
Pritam Banerjee, Suvankar Paul, Tapobrata Sarkar
\thanks{\noindent E-mail:~ bpritam, svnkr, tapo@iitk.ac.in}
\vskip0.4cm
{\sl Department of Physics, \\
Indian Institute of Technology,\\
Kanpur 208016, \\
India}}
\maketitle\abstract{
\noindent
Bogoliubov quasiparticles moving in the background of superfluid $\text{He}^3$-A see an apparently 
curved space-time metric when the background superfluid vacuum is in motion. We study how this curvature couples with the spins 
of the effectively massless quasiparticles. First, we set up the problem in null Fermi coordinates for 
radial as well as circular geodesics and then use it in the context of the analog metric seen by the quasiparticles. 
We obtain an effective magnetic interaction due to curvature coupling, and provide numerical estimates. Some possible 
implications of these results are then pointed out.}
\end{titlepage}
%
%

\section{Introduction and motivation}

The purpose of this paper is to study general relativistic effects of space-time curvature on fermionic spins, via an analog
gravity black hole setup in the context of superfluid $\text{He}^3$-A. In particular, we will focus on such relativistic effects on 
fermionic quasiparticles in the latter system, which are effectively
massless in the background of an analog gravity metric. These quasiparticles move with finite speed, 
which makes this analysis interesting, and we believe that this should have experimental significance. The motivation for this work 
comes from the fact that such analysis (in real black hole scenarios) involving curved space-times are difficult to 
envisage, especially in the context of massless fermions, which would then move with the speed of light. The analog gravity picture
on the other hand offers a somewhat simpler situation to consider, from which useful physical insights can be gleaned. 

The relativistic computations here are carried out in coordinates that are locally flat along a geodesic
trajectory. These are called Fermi normal coordinates. In general, spinning particles do not follow geodesic trajectories,
but the deviation from the latter are known to be small. Since our computations are relevant for effectively massless
fermions, we use null Fermi coordinates, and interpret the results as those that will be seen by a null observer
who moves along a (null) geodesic trajectory, and performs a measurement on the massless fermions. 

In this work, we analyze the interaction Lagrangian for massless fermions in the background of an analog metric in null Fermi
coordinates, and obtain an interaction term (the curvature coupling) via an {\it effective} magnetic field
for these fermions that arise out of analog gravity effects. Standard analysis
in quantum mechanics then implies interesting effects that should arise due to this coupling. 
In order to obtain numerical estimates of our results, we use the uncertainty relation and an energy condition as applicable 
to low energy quasiparticles in  $\text{He}^3$-A. 

The discussion above provides the motivation and methodology of the present work. Similar computations on effective
magnetic field interactions with massive fermions involving Fermi normal coordinates are well known in the literature. However, to 
the best of our knowledge, massless fermions were not analysed in a similar framework, as this will usually not have much physical 
relevance. We claim however that such analysis becomes important and interesting in analog gravity, in the context of 
effectively massless fermionic quasiparticles. This is the main idea developed in this paper. 

In the rest of this paper, we will analyse analog gravity coupling to massless fermionic quasiparticles in the context of an acoustic
black hole. In the next section, we review the necessary background material, to set up the problem. 
Next, in section 3, we will discuss a possible scenario in which such a black hole can be realized in the
laboratory. In section 4, we will first review the known physics of curvature coupling to gravity for massive particles, to set the
notation and conventions. Next, we construct the relevant formulas for such couplings for massless particles, which involves 
determining the null Fermi coordinates. Section 5 deals with curvature couplings of massless fermions that travel in radial
null geodesics, in the context of the black hole described in section 3. In section 6, we present similar results for null circular 
geodesics. In section 7, we provide numerical values that should be useful in an experimental setup. This is done by various
estimates of relevant physical quantities. Section 8 concludes this work with a summary of the results and a 
limitation of our analysis. 

\section{Overview and setup}

Our interest in this paper is in analog gravity, which is known to mimic general relativistic effects, and is an useful tool in
probing effects of gravity in table-top laboratory experiments. We will first briefly review this connection. 

The general theory of relativity (GR) \cite{Weinberg},\cite{Wald},\cite{Hartle},\cite{7} is the most successful theory of gravity till date. 
Several experimental tests of GR have been performed over
the last century, and validation of theoretical predictions in all cases have put the theory on a very firm footing. An interesting arena of 
study has been the effect of gravity on (classical) spins. In this context, the geodetic (or de Sitter) effect and the Lense-Thirring frame 
dragging effect are well known, and the recent gravity probe B experiment \cite{GPB} confirms GR predictions 
on the same to a very high degree of accuracy. 

It is by now an established fact that one can mimic GR effects in condensed matter systems, the topic now being known as analog gravity. 
The importance of this subject lies in the fact that analog systems are relatively easier to probe than 
space-times with curvature, and may provide an easier route to detect gravitational effects in the laboratory, the ultimate aim
being to understand possible quantum effects of gravity, via this indirect route. 
The analog mapping of gravitational effects in condensed matter systems originated from the celebrated 
work of Unruh in 1981 \cite{1},\cite{2}, where it was shown that motion of elastic perturbations (like sound waves) in the 
background of normal moving liquids obey the same equations as that of relativistic massless scalar fields moving in 
a $(3+1)$-dimensional curved space-time, the curvature being determined by an analog metric. 

It was then realized that mimicking gravity with normal liquids might be somewhat difficult, as such liquids will 
contain dissipative effects, that might destroy quantum effects related to the event horizon of the associated black hole. 
Therefore, in order to probe such quantum effects, one should more appropriately look at superfluids. It was then found \cite{3}
that low-energy Bogoliubov quasiparticles in moving superfluid $\text{He}^3 $-A obey similar relativistic equations as found by Unruh.
The literature on analog gravity is rather large, and the subject continues to attract great interest in the community. Various 
effects like Hawking radiation etc. have been extensively investigated in this context, via several pioneering works. 
For a relatively recent review, we refer the reader to the excellent review of \cite{analog1}. 
For more recent works, see, 
e.g, \cite{analog2},\cite{analog3},\cite{analog4},\cite{analog5},\cite{analog6},\cite{analog7},\cite{ParthaDa}.  
In this paper, we consider massless fermions in static metrics, and compute the interaction of gravity with the fermion spin, in 
null Fermi coordinates. We believe that this is novel, and complements the results in the existing literature. 
The motivation for this work largely comes from considerations of analog gravity, and we consider a particular analog black hole
setup involving superfluid $\text{He}^3 $-A. 

In textbook examples, when one considers spin, one usually refers to the vector spin of a gyroscope, or to the expectation value of 
a quantum spin state. However, coupling of gravity to (say) Dirac fermions assumes importance in the study of the hitherto less known 
relation between gravity and 
quantum mechanics. Such studies are also abundant in the literature. For example, the energy spectrum of the Hydrogen atom in a curved 
background was worked out in details by Parker \cite{11},\cite{Parker2},\cite{12}, 
about three decades back, and several similar works have since followed. These are particularly important in the study of possible
Lorentz and CPT violating gravitational interactions. 

In a previous paper \cite{13}, we studied massive Dirac fermions in the background of a generic static space-time. It is well
known (see, e.g \cite{8},\cite{9}) that in such a situation, the interaction Lagrangian in a suitable limit assumes the
form ${\vec B}.{\vec S}$ where ${\vec B}$ is an {\it effective} magnetic field introduced by gravitational interactions. This computation
was done for observers in both radial and circular geodesics, (in Fermi normal coordinates \cite{6}, see below) and it was shown in
\cite{13} that the effective magnetic field can be large, within experimentally measurable ranges for a class of static backgrounds. 

The role of the observer is important in this context, and needs careful explanation. It is known (see, e.g \cite{YeeBander} and
references therein) that in general objects with spin do not follow geodesic paths (although the difference between the actual path of
such objects and corresponding geodesics are small), and the same is expected to hold for 
quantum spins as well. For massive fermions, one has to therefore imagine a hypothetical timelike observer who moves along a geodesic, 
and does measurements on a fermionic system in her reference frame. This observer has to set up a locally flat coordinate system
all along the geodesic, and such a system is constructed by the prescription of Manasse and Misner \cite{6}, and is called the 
Ferm normal coordinate system. For example, in \cite{11}, it was assumed that the 
Hydrogen atom as a whole moves on a geodesic to a good degree of approximation, and the Dirac equation of the electron 
was solved in the Fermi normal system corresponding to a Schwarzschild background, yielding corrections due to gravity in the
energy spectrum. 

Our computations in this paper will involve massless (Weyl) fermions in $\text{He}^3 $-A, and the issue of the observer is 
even more subtle. Indeed,
it is difficult to visualize a null observer (moving at the speed of light) in any physical sense, and although null Fermi coordinates 
can be set up, in general the issue might seem to be purely of mathematical interest. However, there is one situation in which this 
might make physical sense, and this is in the context of analog gravity mentioned above. 
As we discuss in details, in this situation, the speed of
light is replaced by the supercritical quasiparticle speed, and it is not difficult to imagine a null observer in analog gravity 
making a measurement on
a massless fermionic system. Although we are unable to offer a precise experimental setup, we believe that this might be an 
interesting issue to pursue, and it gives rise to some novel effects in analog gravity. We proceed with this understanding. 

In the context of analog gravity in $\text{He}^3$-A,
the quasi-particle spectrum, expanded near some special points on the Fermi surface reduces to that of charged, massless
fermions, with the degrees of freedom propagating on a curved background. It is this physical fact that motivates our analysis in the
present paper. In particular, we will consider the Weyl Lagrangian in the background of the curved metric that the quasiparticles see,
and determine the curvature coupling of the quasiparticles with the background metric. At this point, let us mention some known
facts that we will remember throughout this work. We will follow the discussion of Volovik \cite{VolovikBookHelium} (see section
4.3.2 of that book). 

In the present situation, we deal with particles and (fermionic) quasiparticles. Here, following \cite{VolovikBookHelium}, we can
define two distinct types of observers, the ``external'' observer who deals with the realm of particles and the ``inner'' observer that deals 
with quasiparticles. While the former is not affected by the analog metric, the latter lives in an effective curved space-time that is the 
result of this metric, and free quasiparticles will move in roughly geodesic paths of this metric, according to the inner observer. For the analog
metric that determines the dynamics of massless fermionic quasiparticles in $\text{He}^3 $-A, it is interesting to understand how
the (charged) Weyl fermionic quasiparticles couple to the curvature of the analog metric. Studying this aspect of analog gravity is the
purpose of this paper, and we will see later, might be of relevance to analog gravity experiments. 
To the best of our knowledge, such an analysis has not been performed previously in the literature. 

Once such a framework is set up, we can ask the following question : suppose we consider an external observer (in the sense of
the last paragraph) moving with speed $c$ (the analog of the speed of light, see below) 
in $\text{He}^3 $-A, along a radial or circular path, mimicking the geodesic 
trajectory of the inner observer. The inner observer analyzes the system in her locally flat coordinate system along the geodesic. 
So, if the external observer performs an experiment on the Bogoliubov quasiparticles, we expect her to make 
similar observation as will the inner one. If there is any novel physics that is recorded by the latter, the same should show up
in the observations of our external observer as well. 

To put the above discussion in perspective, we note that 
if the background fluid motion is radial and spherically symmetric, 
the corresponding effective metric is expressed as:
\begin{equation}
ds^2=-\left(c^2-v^2(r)\right)dt^2 \mp 2 v(r) dr dt + dr^2 + r^2 d\Omega^2 \label{a}
\end{equation}
where $ v(r)$ is the velocity of the fluid, and $c$ is the analog of the speed of light.\footnote{We will denote by $c_L 
= 3 \times 10^8~{\rm m/s}$ as the usual speed of light.}
The negative and positive signs above stand for $v(r)>0$ (fluid moves away from center)  and $v(r)<0$ (fluid moves towards the center),
respectively. Also, $d\Omega^2$ is the standard metric on the two-sphere of unit radius. 
If the fluid moves towards the center with increasing velocity and the perturbations move within the fluid with a constant speed $c$, 
the fluid velocity, for a finite radius $ r=r_h $, may equal $c$. This radius will then, mark the position of an event horizon because any 
perturbation moving inside $r_h$ can never come out. Therefore, an analog Black Hole will be formed which Unruh named 
as Sonic Black Hole, and it is expected that many important properties of the black holes can be studied experimentally in the laboratory,
via this analogy. In particular, we will understand the effect of curvature on massless quasiparticles in superfluid backgrounds, which
assumes relevance following our discussion in the last paragraph.


\section{Analog gravity and a possible scenario of a 2-D black hole formation}

In this section, we will briefly review the formalism of analog gravity that will be used in the rest of the paper. 
This section is review material, and we will be brief here. The necessary details can be found in \cite{2},\cite{3},
\cite{VolovikBookHelium},\cite{5}
and references therein. We begin with a brief overview of fermionic quasiparticles in $\text{He}^3 $-A. 

\subsection{Analog gravity in $\text{He}^3 $-A : overview}\label{overview}

It is well known that the energy spectrum of fermionic quasiparticles in $\text{He}^3 $-A is given by 
\begin{equation}
E\left({\vec p}\right) = \pm \sqrt{v_F^2\left(p-p_F\right)^2 + \frac{\Delta_v^2}{p_F^2}\left({\hat l} \times {\vec p}\right)^2}
\label{disp}
\end{equation}
Here, ${\hat l}$ is a unit vector that is directed along the spontaneous angular momentum of the Cooper pairs in the 
$\text{He}^3 $-A condensate, $p_F$ is the Fermi momentum, and given in terms of the Fermi
velocity $v_F$ as $p_F = v_F m^*$, with $m^*$ being an
effective mass, of the order of the mass of the $\text{He}^3$ atom. $\Delta_v$ is called the gap amplitude. 
Zeros of the energy $E({\vec p})=0$ occur at ${\hat l} \times {\vec p} = 0$ and $p=p_F$, which translates into
${\vec p} = ep_F{\hat l}$, with $e=\pm 1$. The quasiparticle energy thus vanishes at ${\vec p} = e{\vec A}$, with 
an effective gauge field ${\vec A} = p_F {\hat l}$.

Low energy excitations can be obtained by expanding the dispersion relation in Eq.(\ref{disp}) in powers of $\left({\vec p} - e{\vec A}\right)$. 
It is not difficult to see that doing this, one obtains (see, e.g \cite{3}) in terms of $c = \Delta_v/p_F$, close to ${\vec p} = e{\vec A}$,
\begin{eqnarray}
&&g^{\mu\nu}\left(p_{\mu} - eA_{\mu}\right)\left(p_{\nu} - eA_{\nu}\right) =0~;~~g^{00}=-1,~\nonumber\\
&&g^{0i}=0,~g^{ij}= v_F^2l^il^j + c^2\left(\delta^{ij} - l^il^j\right)
\label{met}
\end{eqnarray}

The above metric components and the vector potential were specified in terms of the coordinates $\left(x^0,{\vec x}\right)$, where
$x^0$ is the ``Newtonian'' time $t$, and ${\vec x}$ are Cartesian spatial coordinates at rest with respect to the superfluid. However,
for reasons that we describe now, we will be more interested in a situation where the background superfluid is in motion, as it can
give rise to a possible analog black hole. 

\subsection{Analog black hole in $\text{He}^3 $-A}\label{bh}

\begin{figure}[h]
\centering
\includegraphics[height=5 cm,width= 8.5 cm]{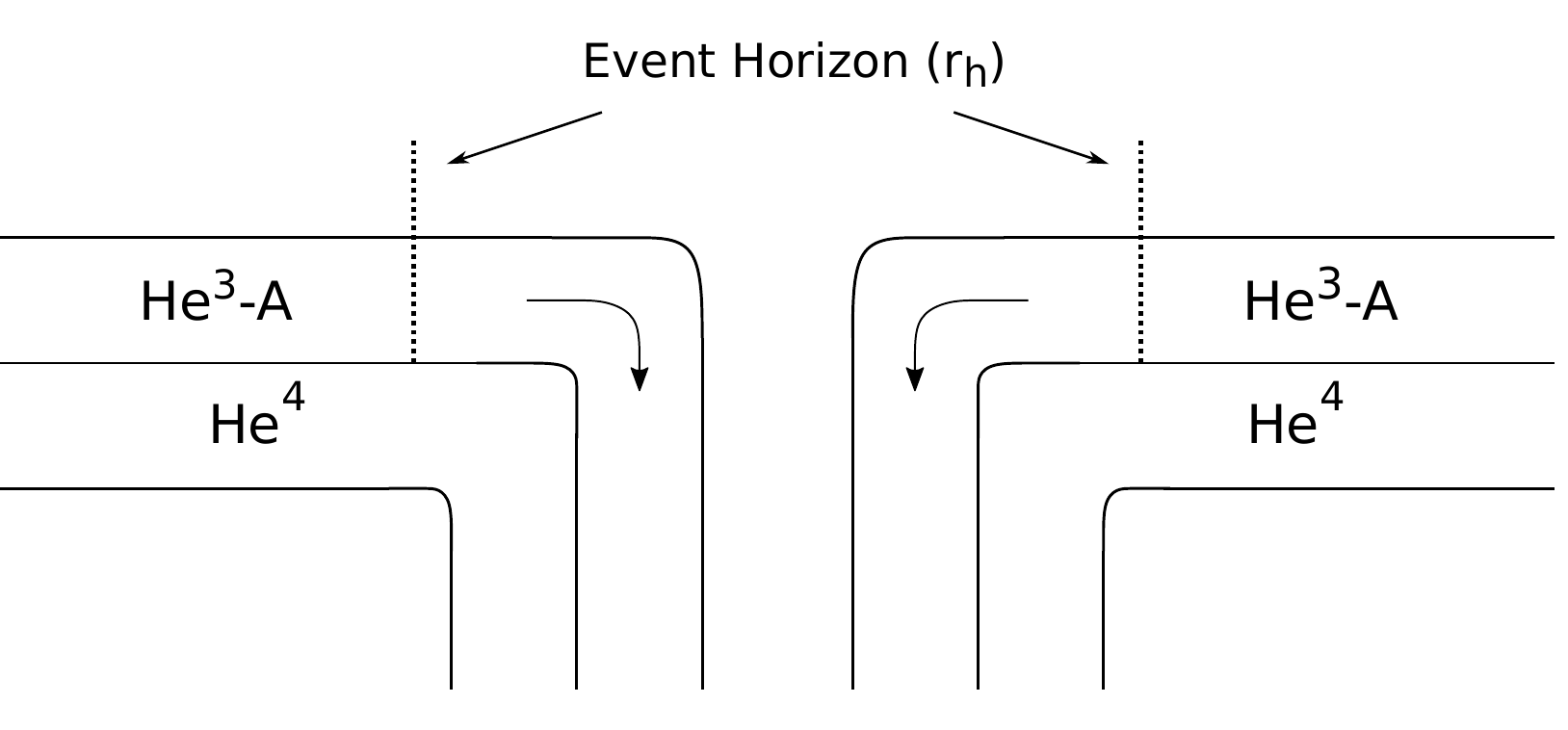}
\caption{Formation of a 2-D black hole in draining bathtub geometry. This is inspired by Fig.1(a) of \cite{5}.} \label{fg1}
\end{figure}

We will be interested in an analog black hole in the context of $\text{He}^3 $-A.
In fact, a two dimensional analog Black Hole formation can be realized in the laboratory if we consider the geometry as described by Volovik, 
in Fig.1(a) of his paper \cite{5} (see Fig.\ref{fg1}), which is also a generalization of the original draining bathtub geometry 
discussed in \cite{4}.

In this figure, a 2-D thin film of $\text{He}^3$-A, which forms the background of the system, is moving towards the orifice at the 
center where it goes into the third (vertical) dimension. But it is not directly connected to the wall of the container. 
The $\text{He}^3$-A film is moving over a thin layer of superfluid $ \text{He}^4 $ as shown. The reason of introducing
superfluid $ \text{He}^4 $ film lies in the fact that the direct motion of $\text{He}^3$-A with respect to the wall of the container
gives rise to some undesirable effects. Namely, the interaction of $\text{He}^3$-A with the walls produces the Cherenkov radiation 
of quasiparticles which collapses the superfluidity and as a result, the `supercritical' motion becomes unstable \cite{5}. 
But the supercritical velocity ($c\sim 3$ cm/s) of $\text{He}^3$-A is much smaller than the Landau velocity for quasiparticle 
radiation in superfluid $ \text{He}^4 $ (about 50 cm/s). So the $ \text{He}^4 $ layer is not excited even when $\text{He}^3$-A moves 
with velocity greater than $c$. This makes the super-critical motion of $\text{He}^3$-A film with respect to $ \text{He}^4 $ layer stable.

To describe the geometry, it is convenient to choose a cylindrical coordinate system. The plane of the $ \text{He}^3 $ film 
is described by the coordinates $r$ and $\phi$. The $z$ axis is set to be normal to the film. If the thickness of the film 
remains constant throughout, 
the velocity of flow of superfluid $\text{He}^3$-A towards the center  will be given as, $v(r)=-\frac{c r_h}{r}$. As it reaches the event 
horizon, $r=r_h$, its velocity equals $c$. Below $r_h$, it is greater than $c$ and above, less than $c$.
As a result, the quasiparticles can never come out of the region  $r< r_h$, forming a Black Hole type scenario.

\subsection{The analog black hole metric}

With the above discussion, it is now possible to write down the metric that is seen by Bogoliubov quasiparticles in 
$\text{He}^3 $-A. Clearly, the formula in Eq.(\ref{met}) of subsection (\ref{overview}) needs to be modified when the background
superfluid $\text{He}^3 $-A is in motion. This is obtained by a simple coordinate transformation of the coordinates used 
in Eq.(\ref{met}), as $\left(t,{\vec x}\right) \to \left(t, {\vec x} - {\vec v}t\right)$. One chooses a coordinate system in which
${\hat l}$ (also the anisotropy axis of the velocity field $v$) points in the direction of ${\hat z}$, and after writing the energy dispertion 
relation of Eq.(\ref{disp}) in the new coordinate system, it can be readily verified \cite{5} that 
the energy spectrum for the low-energy Bogoliubov fermionic quasiparticles yields
\begin{equation}
(E-\mathbf{p}\cdot \mathbf{v})^2 = c^2 (p^2_x + p^2_y) + v^2_F (p_z -e p_F)^2 \label{b}
\end{equation}
where $e= \pm 1$. The velocity of the quasiparticles along the film ($c\sim 3$ cm/s) is 
much smaller than the velocity normal to the film ($v_F \sim 55$ m/s). So the degree of anisotropy of the velocity is large. Again, the 
velocity of the background superfluid vacuum ($v$) is purely 2-dimensional outside the orifice, i.e., 
it moves in the $ (r,\phi) $-plane only. 
Considering such a velocity field, the energy spectrum of the Bogoliubov quasiparticles in Eq.(\ref{b}) can be recast into an effective 
motion of a charged, massless relativistic particle in a $ (3+1) $-dimensional curved space-time with the following form of the metric
\begin{eqnarray}
ds^2 = - \left(c^2 - v^2(r)\right) dt^2 + 2 v(r) drdt + dr^2 + r^2 d\phi^2 + \frac{c^2}{v^2_F} dz^2 \label{c}
\end{eqnarray}
where we have denoted ${\vec v} = \left(v_x, v_y\right)$. 
The above line element also shows that the $ g_{00} $ component of the metric changes sign as $ v(r) $ 
becomes greater than $ c $ inside $ r=r_h $ confirming the formation of a black hole. 

We mention in passing that one can, in this case, make a coordinate transformation 
\begin{equation}
t = \tau + \int \frac{v(r)dr}{c^2 - v(r)^2}
\end{equation}
This equation is integrable, and in terms of the coordinate $\tau$, yields the metric 
\begin{equation}
ds^2 = -\left(c^2 - v^2\right)d\tau^2 + \frac{c^2}{c^2 - v^2}dr^2 + r^2d\phi^2 + \frac{c^2}{v_F^2}dz^2
\label{metfinal}
\end{equation}
The metrics in Eq.(\ref{c}) or (\ref{metfinal}) give equivalent results. We will use the form in Eq.(\ref{c}) in what follows. 
As mentioned in the introduction, the quasiparticles see this metric, and their dynamics is governed by the same. Since these
are analogous to massless charged fermions, it is pertinent to ask what a null observer in the background of the metric of
Eq.(\ref{c}) would say about the fermions.

In the remainder of this paper, we study massless fermionic quasiparticles in the background of
the geometry of Eq.(\ref{metfinal}) (equivalently Eq.(\ref{c})). In the next section, we present the formalism to deal with this problem. 


\section{Curvature coupling of quasiparticles : formalism}

Let us imagine that the superfluid excitations, known as Bogoliubov quasiparticles, move in geodesics in an effectively curved space-time. 
Since these quasiparticles are excitations of superfluid $\text{He}^3 $-A vacuum, they are just the dressed $ \text{He}^3 $ atoms having 
Bogoliubov spin, and are fermionic in nature. As a result, they will exhibit the characteristic signatures of their spin while moving in 
(nearly) geodesic trajectories by getting coupled with the intrinsic curvature of the space-time metric that they see. 
As mentioned before, we have in mind an observer who makes a measurement on the fermions, in coordinates that
are locally flat all along a given geodesic. To set the notations and
conventions, let us first discuss the known simpler formalism of computing curvature couplings for massive fermions in brief, and then we 
will apply it in our specific problem.

\subsection{Curvature coupling of massive fermions}

Fermions in curved space-time is a well researched topic. In this paper, we are interested in the Fermi normal coordinates, i.e 
a coordinate system that is locally flat at each point of space-time through which the spinor travels. An extensive discussion and construction 
of Fermi Normal Coordinates for timelike geodesics can be found in \cite{6} by Manasse and Misner. Let us briefly review the
construction of \cite{6}, to set the stage.  

Consider a set of four orthogonal vectors which satisfies the following two relations along a timelike geodesic of a massive particle \cite{6}
\begin{equation}
\hat{e}_\alpha \cdot \hat{e}_\beta = \eta_{\alpha \beta} ~~ , ~~~~ \nabla_{\nu'}(\hat{e}^{\mu'}_\alpha) \hat{e}^{\nu'}_0 = 0 \label{d}
\end{equation}
where $ \nabla $ denotes covariant derivative, $ \eta_{\alpha \beta} $ is the usual Minkowski metric with signature $ (-,+,+,+) $ and 
$ \hat{e}_0 $ represents the tangent vector to the timelike geodesic. The primed indices refer to the components of the vectors in the 
original coordinate system of the metric, and the unprimed indices refer to the corresponding components in Fermi normal coordinates. 
The structures $ \hat{e}^{\mu'}_\alpha,\hat{e}^{\nu'}_\beta... $ define the different elements of the coordinate transformation matrix from 
general coordinates to Fermi normal coordinates. Therefore, once the above tetrad is set as the basis of Fermi normal coordinate system, 
we can in principle compute the components of every tensor in this locally flat system. For Riemann curvature tensor, these components are
\begin{equation}
R_{\alpha \beta \gamma \delta} = \hat{e}^{\mu'}_\alpha \hat{e}^{\nu'}_\beta \hat{e}^{\lambda'}_\gamma \hat{e}^{\sigma'}_\delta R_{\mu' \nu' \lambda' \sigma'} \label{e}
\end{equation}
The metric close to the geodesic $ (G) $, now, looks like, up to second order in coordinates \cite{6,7}
\begin{eqnarray}
g_{00} &=& -1 - R_{0l0m}\vert_G ~ x^l x^m, ~~ g_{0i} = -\frac{2}{3} R_{0lim}\vert_G ~ x^l x^m \nonumber\\
g_{ij} &=& \delta_{ij} - \frac{1}{3} R_{iljm}\vert_G ~ x^l x^m \label{f}
\end{eqnarray}
where the Latin indices $ i,j,k,... $ take the values 1,2 and 3. Here, the observer's time dependence enters the metric only through 
the curvature tensor components as they are evaluated at a particular proper time along the geodesic $ G $. After obtaining such a 
coordinate system, we can study the covariant Dirac Lagrangian given by
\begin{equation}
\mathcal{L} = \sqrt{-g}(i \bar{\psi}\gamma^\alpha \mathcal{D}_\alpha \psi - m \bar{\psi}\psi) \label{g}
\end{equation}
where $ \gamma^\alpha $ are the normal Dirac matrices.

The definition of $ \mathcal{D}_\alpha $ (covariant derivative) in Eq.(\ref{g}) is given by
\begin{equation}
\mathcal{D}_\alpha \equiv (\partial_\alpha -\frac{i}{4}\omega_{\beta \gamma \alpha} \sigma^{\beta \gamma}) \label{h}
\end{equation}
where the spin connection $ (\omega_{\beta \gamma \alpha}) $ and $ \sigma^{\beta \gamma} $ are given, respectively, by
\begin{eqnarray}
\omega_{\beta \gamma \alpha} &=& \hat{e}_{\beta \mu'}(\partial_\alpha \hat{e}^{\mu'} _\gamma 
+ \Gamma^{\mu'} _{\nu' \rho'} \hat{e}^{\nu'} _\gamma \hat{e}^{\rho'} _\alpha),\nonumber\\
\sigma^{\beta \gamma} &=& \frac{i}{2}[\gamma^\beta , \gamma^\gamma] \label{i}
\end{eqnarray}
In the above expressions, $ \Gamma^{\mu'} _{\nu' \rho'} $ are the Christoffel connection and $ e^{\mu'} _\alpha $ denotes, as stated before, 
the coefficient of the transformation matrix connecting the curved and flat space-times. If the expression of $ \mathcal{D}_\alpha $ is 
put in the Lagrangian equation, i.e, Eq.(\ref{g}), the corresponding term coming from the spin connection involves an interaction Lagrangian 
of the form $ \bar{\psi}\gamma^\alpha \gamma^5 b_\alpha \psi $ \cite{8,9}. The four vector $ b^\alpha $ can be written in the following form :

\begin{eqnarray}
b^\sigma &=& \frac{1}{4}\epsilon ^{\alpha \beta \gamma \sigma}\hat{e}_{\beta \mu'}(\partial_\alpha \hat{e}^{\mu'}_\gamma + 
\Gamma^{\mu'} _{\nu' \rho'} \hat{e}^{\nu'}_\gamma \hat{e}^{\rho'}_\alpha) \nonumber \\
&\equiv& \frac{1}{4}\epsilon ^{\alpha \beta \gamma \sigma}\hat{e}_{\beta \mu'} \partial_\alpha \hat{e}^{\mu'}_\gamma \label{j}
\end{eqnarray}
where $ \gamma^5 = i\gamma^0 \gamma^1 \gamma^2 \gamma^3 $. In the corresponding Hamiltonian, this interaction term can be 
made to look like an effective interaction energy of the form $ -\vec{b} \cdot \vec{s} $ in the non-relativistic limit \cite{10}. Here, 
both $ \vec{b} $ and $ \vec{s} $ represent normal 3-dimensional vectors as $ b_0 $ is identically zero according to Eq.(\ref{j}).

Now, to proceed further in this analysis, we need to find out the expressions of the effective magnetic field components ($ b^\sigma $) in 
Fermi normal coordinates. The forms of the metric components in these coordinates are already given in Eq.(\ref{f}) and the corresponding 
expressions of the vierbeins are \cite{11,12}
\begin{eqnarray}
&&\hat{e}^{\mu'} _0 = \delta^{\mu'}_0 -\frac{1}{2}R^{\mu'}{}_{l0m}\vert_G ~ x^l x^m, \nonumber\\
&&\hat{e}^{\mu'} _i = \delta^{\mu'}_i -\frac{1}{6}R^{\mu'}{}_{lim}\vert_G ~ x^l x^m \label{k}
\end{eqnarray}
where $ i,l,m,... $ run over spatial indices only. Therefore, we shall first set up the tetrad basis of Eq.(\ref{d}) and calculate the components 
of the Riemann tensor in these coordinates using Eq.(\ref{e}). Then we put the expressions of vierbeins of Eq.(\ref{k}) into Eq.(\ref{j}) 
to obtain the desired result in Fermi normal coordinates. The corresponding forms of $ b_\sigma $ come out to be 
\begin{eqnarray}
b_0 &=& -\frac{1}{4}\epsilon_{0ijk}\left(\frac{2}{3}R^{jik}{}_m\big\rvert_G +\frac{1}{6}R^{jki}{}_m\big\rvert_G\right) ~ x^m, \nonumber \\
b_i &=& \frac{1}{4}\epsilon_{0ijk}\left(\frac{1}{3}R^0{}_{m}{}^{jk}\big\rvert_G  - \frac{1}{3}R^{0k}{}_m{}^j \big\rvert_G \right)x^m \label{l}
\end{eqnarray}
The above expression of $ b_0 $ is such that it vanishes identically, as stated before. After determining 
$ R_{\alpha \beta \gamma \delta} $ in Fermi normal coordinates from Eq.(\ref{e}), we can straightforwardly find out the effective 
magnetic filed ($ \vec{b} $) due to gravitational effects in the non-relativistic limit. Although the term `magnetic field' is used here 
for its $ -\vec{b}\cdot \vec{s} $ type contribution to the Hamiltonian, it is different from any intrinsic magnetic field that may be present 
in the system. Moreover, it is to be noted that the measurement of this magnetic field is done in the vicinity of the geodesic as exactly 
on the geodesic the $ x^l $ are zero, forcing $ \vec{b} $ to vanish identically.


\subsection{Curvature coupling of massless particles and Null Fermi coordinates}

The entire analysis described in the previous subsection is done with the basic assumption that the moving particle is massive and thus the 
corresponding geodesic is timelike. But in analog gravity of $ \text{He}^3$-A, the quasiparticles are in principle massless fermions moving 
in a curved space-time as described earlier. Therefore, it is more useful to consider a null observer, and 
we need to reformulate the above analysis for null geodesics, and find the modified 
expressions of curvature couplings. The first question that arises in this regard is how to define the notion of Fermi normal coordinates 
for null geodesics. The construction here is somewhat subtle, and has been recently addressed in \cite{16}. The technical subtlety here 
is that since the tangent to a null geodesic is a null vector, the corresponding set of four vectors which act as the basis of null Fermi 
coordinates cannot be orthonormal.

Following the construction of \cite{16}, let us define four pseudo-orthonormal vectors satisfying the same two relations as given 
in Eq.(\ref{d}) along a null geodesic $ \mathcal{N} $
\begin{equation}
\hat{E}_A \cdot \hat{E}_B = \eta_{AB} ~~ , ~~~~ \nabla_{\nu'}(\hat{E}^{\mu'}_A) \hat{E}^{\nu'}_+ = 0 \label{w}
\end{equation}
where $ \hat{E}_+ $ is tangent to the null geodesic and $ \eta_{AB} $ is still the flat Minkowski metric but expressed in a new 
$ E^A $-basis. The matrix form of $ \eta_{AB} $ in this new basis and the corresponding line element along $ \mathcal{N} $ 
are given by
\begin{equation}
	\eta_{AB} = \left( {\begin{array}{cccc}
		                          0 & 1 & 0 & 0 \\
		                          1 & 0 & 0 & 0 \\
		                          0 & 0 & 1 & 0 \\
		                          0 & 0 & 0 & 1 \\
		                          \end{array} } \right) ,~ 
	ds^2 \vert_\mathcal{N} = 2 E^+ E^- + \delta_{ab} E^a E^b
\label{etaform}
\end{equation}
where each $ A,B,... $ takes the values $ (+,-,2,3) $ and each $ a,b,... $ takes $ (2,3) $. The corresponding Fermi coordinates 
of a point $ x $ are denoted as $ (x^A)=(x^+, x^-, x^a) $ and their definition is given in \cite{16}.
Once again, the quantities $ \hat{E}^{\mu'}_A $ represent the different elements of the basis transformation matrix from the 
actual $ x^{\mu'} $ coordinate system to the Fermi coordinate system $ x^A $. Along $ \mathcal{N} $, the two coordinate 
systems are related by
\begin{equation}
\frac{\partial{x}^A}{\partial{x}^{\mu'}} \biggr \rvert_\mathcal{N} = \hat{E}^A_{\mu'}~, 
~~~ \frac{\partial{x}^{\mu'}}{\partial{x}^A} \biggr \rvert_\mathcal{N} = \hat{E}^{\mu'}_A \label{x}
\end{equation}

The components of Riemann curvature tensor in Fermi coordinates are evaluated from the equation which is similar to Eq.(\ref{e}), and
are given by
\begin{equation}
R_{ABCD} = \hat{E}^{\mu'}_A \hat{E}^{\nu'}_B \hat{E}^{\lambda'}_C \hat{E}^{\sigma'}_D R_{\mu' \nu' \lambda' \sigma'} \label{y}
\end{equation}
The components of the metric tensor in the vicinity of the geodesic $ \mathcal{N} $, up to second order, can be
shown to be given by
\begin{eqnarray}
g_{++} &=& -R_{+ \bar{c} + \bar{d}} ~ \big \rvert_\mathcal{N} ~~ x^{\bar{c}} x^{\bar{d}} ~, ~~
g_{--} = -\frac{1}{3} R_{- \bar{c} - \bar{d}} ~~ \big \rvert_\mathcal{N} ~ x^{\bar{c}} x^{\bar{d}} ~,\nonumber\\
g_{+-} &=& 1 -\frac{2}{3} R_{+ \bar{c} - \bar{d}} ~ \big \rvert_\mathcal{N} ~~ x^{\bar{c}} x^{\bar{d}} ~, ~~
g_{ab} = \delta_{ab} -\frac{1}{3} R_{a \bar{c} b \bar{d}} ~ \big \rvert_\mathcal{N} ~~ x^{\bar{c}} x^{\bar{d}} ~~ ,\nonumber\\
g_{+a} &=& -\frac{2}{3} R_{+ \bar{c} a \bar{d}} ~ \big \rvert_\mathcal{N} ~~ x^{\bar{c}} x^{\bar{d}} ~, ~~
g_{-a} = -\frac{1}{3} R_{- \bar{c} a \bar{d}} ~ \big \rvert_\mathcal{N} ~~ x^{\bar{c}} x^{\bar{d}} \label{z}
\end{eqnarray}
where $ (\bar{a})= (-,a) $.

The covariant Dirac Lagrangian for massless fermions is
\begin{equation}
\mathcal{L} = i \sqrt{-g} \bar{\psi}\gamma^A \mathcal{D}_A \psi \label{aa}
\end{equation}
where $ \mathcal{D}_A $ is given by Eq.(\ref{h}) (apart from a term involving an effective gauge field), 
with $ \alpha $'s replaced by $ A $'s, and the corresponding expressions of spin connection and $ \sigma^{AB} $ 
are also similar to Eq.(\ref{i}) :
\begin{eqnarray}
\omega_{BCA} &=& \hat{E}_{B \mu'}(\partial_A \hat{E}^{\mu'} _C + \Gamma^{\mu'} _{\nu' \rho'} \hat{E}^{\nu'} _C \hat{E}^{\rho'} _A),\nonumber\\
\sigma^{BA} &=& \frac{i}{2}[\gamma^B , \gamma^A] \label{ab}
\end{eqnarray}
We note here that there is an extra term in the covariant derivative, involving the effective gauge field, as follows
from the discussion of subsection 3.1 (see \cite{VolovikWeyl}). Inclusion of this additional term makes the expressions cumbersome, 
and for the moment we will work with the terms of Eq.(\ref{h}) purely for ease of presentation, and the term involving the gauge field 
will be introduced later, following Eq.(\ref{fullCovDer}). 

Here, we will have to be careful in defining $ \gamma^A $. Unlike the previous case where each $ \gamma^\alpha $ 
represents one of the standard Dirac matrices, the forms of $ \gamma^A $'s, in this case are different. 
Note that the Lagrangian in flat space-time for massless fermions can be decomposed into two parts
\begin{eqnarray}
\mathcal{L'} &=& i \bar{\psi}\gamma^{\mu} \partial_{\mu} \psi  \nonumber \\
&=& i u_{-}^\dagger \sigma^{\mu} \partial_{\mu} u_{-} + i u_{+}^\dagger \bar{\sigma}^{\mu} \partial_{\mu} u_{+} \label{ac}
\end{eqnarray}
where $ \sigma^{\mu} = (1, \sigma^i ) $, $ \bar{\sigma}^{\mu} = (1, -\sigma^i ) $ with $ \sigma^i $'s being the Pauli matrices and 
$\psi = \left(u_+,u_-\right)^T$. 
In case of a massive fermion, $ u_{+} $ and $ u_{-} $ cannot be separated completely, but we can describe a massless fermion 
by $ u_{+} $ or $ u_{-} $ alone with the respective equation of motion given by
\begin{equation}
i \bar{\sigma}^{\mu} \partial_{\mu} u_{+} = 0,~~\text{or} ~~ i \sigma^{\mu} \partial_{\mu} u_{-} = 0 \label{ad}
\end{equation}
These equations are the well known Weyl equations for massless fermions, and involve Pauli matrices.
Now, let us apply this analysis to the covariant Dirac Lagrangian for massless fermions expressed in null Fermi coordinates, Eq.(\ref{aa})
\begin{eqnarray}
\mathcal{L} = i \sqrt{-g} ~ u_{-}^\dagger \tilde{\sigma}^A \mathcal{D}_A u_{-} + 
i \sqrt{-g} ~ u_{+}^\dagger \bar{\tilde{\sigma}}^A \mathcal{D}_A u_{+} ~\left(\equiv {\mathcal L}_1 +
{\mathcal L}_2\right) \label{ae}
\end{eqnarray}

The corresponding Weyl equations for $ u_{+} $ or $ u_{-} $ will be respectively
\begin{equation}
i \bar{\tilde{\sigma}}^A \mathcal{D}_A u_{+} = 0,~~\text{or} ~~ i \tilde{\sigma}^A \mathcal{D}_A u_{-} = 0 \label{af}
\end{equation}
Let us compare the second expressions of equations (\ref{ad}) and (\ref{af}). These expressions are similar with $ \partial_{\mu} $ 
replaced by $ \mathcal{D}_A $ and $ \sigma^{\mu} $ replaced by $ \tilde{\sigma}^A $. The difference between 
$ \sigma^{\mu} $ and $ \tilde{\sigma}^A $ is easy to understand. $ \sigma^{\mu} $ in Eq.(\ref{ad}) is just the Pauli matrices 
with the background flat metric given by ${\rm Diag}\left(-1,1,1,1\right)$. 
The forms of the $ \tilde{\sigma}^A $ in Eq.(\ref{af}) are different from Pauli matrices. The reason for this is that $ \tilde{\sigma}^A $ 
is expressed in terms of null Fermi coordinates and if we follow the definition (\ref{w}) of pseudo-orthonormal Fermi frames with 
$ \hat{E}^+ $ and $ \hat{E}^- $ being null vectors, the corresponding background locally flat metric along a null geodesic takes the form
given in the first expression of Eq.(\ref{etaform}). 

Therefore, the transformation relations from $ \eta_{\mu \nu} \longrightarrow \eta_{AB} $ have to be applied on $ \sigma^{\mu} $ to 
obtain the corresponding expressions of $ \tilde{\sigma}^A $ in the new coordinate system. The forms of $ \tilde{\sigma}^A $, 
after this transformation, are given as $ \tilde{\sigma}^A = (\tilde{\sigma}^{+}, \tilde{\sigma}^{-}, \tilde{\sigma}^{2}, \tilde{\sigma}^{3} ) $,
where we have defined
\begin{eqnarray}
  \tilde{\sigma}^{+} = -\frac{1}{\sqrt{2}} \sigma^0 + \frac{1}{\sqrt{2}} \sigma^1 = \frac{1}{\sqrt{2}} \left( {\begin{array}{cc}
    -1 & 1 \\ 1 & -1
  \end{array} } \right)~, \nonumber\\
  \tilde{\sigma}^{-} = \frac{1}{\sqrt{2}} \sigma^0 + \frac{1}{\sqrt{2}} \sigma^1 = \frac{1}{\sqrt{2}} \left( {\begin{array}{cc}
1 & 1 \\
1 & 1 
\end{array} } \right) ~ ,\nonumber\\
 \tilde{\sigma}^{2} = \sigma^{2} = \left( {\begin{array}{cc}
0 & -i \\
i & 0 
\end{array} } \right) ,~
\tilde{\sigma}^{3} = \sigma^{3} = \left( {\begin{array}{cc}
1 & 0 \\
0 & -1 
\end{array} } \right)~~
\end{eqnarray}
Since the forms of $\tilde{\sigma}^A$ are changed from the usual Pauli matrices, so do those of the 
corresponding $ (4 \times 4) $ $ \gamma^A $ matrices, and as a result, they do not exactly resemble 
the Dirac matrices. In particular, we will use the following forms of the $\gamma^A$ matrices :
\begin{eqnarray}
\gamma^+ = \begin{pmatrix} 0 & {\tilde \sigma}^+ \\ -{\tilde \sigma}^- & 0\end{pmatrix},~~~
\gamma^- = \begin{pmatrix} 0 & {\tilde \sigma}^- \\ -{\tilde \sigma}^+ & 0\end{pmatrix}~,~~
\gamma^a = \begin{pmatrix} 0 & {\tilde \sigma}^a \\ -{\tilde \sigma}^a & 0\end{pmatrix}~~(a=2,3)
\end{eqnarray} 
With the new definitions and expressions of $ \tilde{\sigma}^A $, we are now in a position to define the 
curvature coupling of massless fermions expressed in null Fermi coordinates. The expressions of vierbeins analogous to Eq.(\ref{k}) are
\begin{eqnarray}
&~&\hat{E}^{A} _{+} = \delta^{A}_{+} -\frac{1}{2}R^{A}{}_{\bar{c} {+} \bar{d}} ~ \vert_{\mathcal{N}} ~ x^{\bar{c}} x^{\bar{d}},~\nonumber\\
&~&\hat{E}^{A} _{-} = \delta^{A}_{-} -\frac{1}{6}R^{A}{}_{\bar{c} {-} \bar{d}} ~ \vert_{\mathcal{N}} ~ x^{\bar{c}} x^{\bar{d}},~\nonumber\\
&~&\hat{E}^{A} _{a} = \delta^{A}_{a} -\frac{1}{6}R^{A}{}_{\bar{c} {a} \bar{d}} ~ \vert_{\mathcal{N}} ~ x^{\bar{c}} x^{\bar{d}} \label{ag}
\end{eqnarray}
The corresponding expressions of affine connections, in Fermi coordinates, are
\begin{eqnarray}
\Gamma^A{}_{B +} ~ \vert_{\mathcal{N}} = R^{A}{}_{B \bar{a} +} \vert_{\mathcal{N}} ~ x^{\bar{a}},~~~
\Gamma^A{}_{\bar{b} \bar{c}} ~ \vert_{\mathcal{N}} = -\frac{1}{3} \left( R^{A}{}_{\bar{b} \bar{c} \bar{d}} + R^{A}{}_{\bar{c} \bar{b} \bar{d}} \right) \vert_{\mathcal{N}} ~ x^{\bar{d}} \label{ah}
\end{eqnarray}

Similar to the previous section, if we expand the first term of the Lagrangian (\ref{ae}) in Fermi coordinates by using the expressions of 
Eq.(\ref{z}), Eq.(\ref{ag}) and Eq.(\ref{ah}), it takes the following form
\begin{equation}
\mathcal{L}_1 = \sqrt{-g} ~ u_-^{\dagger} \left( i \tilde{\sigma}^A \partial_A + b^A \tilde{\sigma}_A + i a^A \tilde{\sigma}_A \right) u_-
\label{Lag1}
\end{equation}
The third term which is anti-hermitian vanishes when its conjugate part is added to the Lagrangian. Therefore, the only interaction term that 
survives is the second one which is hermitian. The expressions of the components of this gravitational coupling term ($ b^A $) come out to be
\begin{eqnarray}
b^{+} &=& \frac{1}{4} \epsilon^{1ab} \left[ \frac{1}{6} \left( R_{+ \bar{m} a b} - R_{- a b \bar{m}} \right)  
+ \frac{1}{3} \left( R_{+ a b \bar{m}} + R_{- b a \bar{m}} \right) + \frac{1}{2} \left( R_{- \bar{m} a b} +
 R_{+ b a \bar{m}} \right) \right] x^{\bar{m}}\nonumber\\
b^{-} &=& -\frac{1}{4} \epsilon^{1ab} \left[ \frac{1}{6} R_{- \bar{m} a b} + \frac{7}{6} R_{+ \bar{m} a b} +
\frac{1}{3} \left( R_{+ a b \bar{m}} + R_{- a b \bar{m}} \right) + \frac{1}{2} \left( R_{- b a \bar{m}} + 
R_{+ b a \bar{m}} \right) \right] x^{\bar{m}}\nonumber\\
b^{c} &=& \frac{1}{4} \epsilon^{1ac} \left[ \frac{7}{6} R_{- a \bar{m} +} - R_{+ a + \bar{m}} + 
\frac{1}{6} R_{- a - \bar{m}} + \frac{1}{3} R_{+ a - \bar{m}} + \frac{1}{2} \left( R_{+ - a \bar{m}} + 
R_{- - a \bar{m}} \right) \right] x^{\bar{m}}~~~~~~~ \label{ai}
\end{eqnarray}
where again $ (\bar{a})=(-,a) $ and $ a,b,c,... $ take values $(2, 3)$. 

The above expressions were evaluated for the form of the covariant derivative given in Eq.(\ref{h}). Including the vector potential term,
the full covariant derivative is given by \cite{VolovikWeyl} : 
\begin{equation}
\mathcal{D}_A \equiv \partial_A - \frac{i}{4} \omega_{BDA} \sigma^{BD} - i \tilde{A}_A
\label{fullCovDer}
\end{equation}
where, $ \tilde{A}_A = A_A + \chi_A $, with the expressions of $ \chi_A $ and $A_A$ being 
\begin{eqnarray}
\chi_A = \frac{1}{8} \epsilon^{\lambda' \gamma' \mu' \nu'} E_{A \lambda'} ~ E^B_{\gamma'}
\left( \partial _{\mu'} E_{B \nu'} - \partial _{\nu'} E_{B \mu'} \right)~,~~
 A_A = \left(0,0,0,p_F \right)
\end{eqnarray}
The corresponding expressions of $ \chi^A $ in Fermi coordinates are evaluated to be 
\begin{eqnarray}\label{Ki}
\chi^{+} &=& 0~, \nonumber\\
\chi^{-} &=& \frac{1}{4} \left( \frac{1}{3} R_{+32\bar{m}} + \frac{2}{3}
R_{+\bar{m}23} - \frac{1}{3} R_{+23\bar{m}} \right) x^{\bar{m}}~, \nonumber\\
\chi^{2} &=& \frac{1}{4} \left( \frac{2}{3} R_{+\bar{m}3-} - \frac{1}{3}
R_{+3-\bar{m}} + \frac{1}{3} R_{+-3\bar{m}} \right) x^{\bar{m}}~, \nonumber\\
\chi^{3} &=& \frac{1}{4} \left( -\frac{2}{3} R_{+\bar{m}2-} + \frac{1}{3}
R_{+2-\bar{m}} - \frac{1}{3} R_{+-2\bar{m}} \right) x^{\bar{m}}
\end{eqnarray}
The total magnetic field including the gauge field term is now given by 
\begin{equation}
B^A = b^A + \chi^A + A^A 
\label{Bfinal}
\end{equation}
with the form of $b^A$ given in Eq.(\ref{ai}). Note that the last term in Eq.(\ref{Bfinal}) is a constant term, 
and we will ignore this in our analysis. In what follows, we will focus on the first two terms of Eq.(\ref{Bfinal}) and 
in the next section, we proceed to evaluate the components of 
$B^A $ for both radial and circular null geodesics in the background of analog gravity and study its characteristic features in some details.

\section{Massless fermionic quasiparticles in radial null geodesics}

We start with the metric of Eq.(\ref{c}) which we reproduce here for convenience : 
\begin{eqnarray}
ds^2 =-\left(c^2 - v^2(r)\right) dt^2 + 2 v(r) drdt + dr^2 
+ r^2 d\phi^2 + \frac{c^2}{v^2_F} dz^2 \nonumber
\end{eqnarray}
For null geodesics in this space-time, the normalization of the four-velocity yields
\begin{equation}
\dot{r}^2 + 2 v(r) \dot{t} \dot{r} + r^2 \dot{\phi}^2 - \left(c^2-v^2(r)\right)\dot{t}^2  + \frac{c^2}{v_F^2}{\dot z}^2 = 0 \label{o}
\end{equation}
where over-dots represent derivative with respect of an affine parameter along the null geodesic. 
For timelike geodesics, a standard choice of this affine parameter is the proper time. But in case of a null geodesic, the 
affine parameter cannot be the proper time. Instead, normal coordinate time or radial distance may be considered as the affine parameter,
if they satisfy the geodesic equation of the form
\begin{equation}
\nabla_{\nu'} (u^{\mu'}) u^{\nu'} = 0 \label{aj}
\end{equation}
where $ u^{\mu'} $ is the tangent vector to the null geodesic under consideration. Here, by 
radial null geodesics, we mean the set of null geodesics for which $ \dot{\phi} = \frac{d\phi}{d\lambda} = 0 $, 
with $ \lambda $ being the affine parameter. Therefore, for radial null geodesics outside the orifice, Eq.(\ref{o}) becomes
\begin{equation}
\dot{r}^2 + 2 v(r) \dot{t} \dot{r} - \left(c^2-v^2(r)\right)\dot{t}^2  = 0 \label{ak}
\end{equation}

With condition (\ref{ak}) in mind, we can find out the pseudo-orthonormal Fermi tetrad basis for the analog metric along a 
null radial geodesic as :
\begin{eqnarray}
\hat{E}^{\mu'}_+ &=& \left( \frac{c+v(r)}{c^2-v(r)^2} , 1 , 0 , 0 \right)\nonumber\\
\hat{E}^{\mu'}_- &=& \left( \frac{v(r)-c}{2 c^2} - \frac{kc + kv(r)}{2 \left( c^2-v^2(r) \right)} , 
\frac{c^2-v(r)^2}{2c^2} - \frac{k}{2} , \frac{k}{r} , 0 \right)\nonumber\\
\hat{E}^{\mu'}_2 &=& \left( -\frac{kc + kv(r)}{c^2-v^2(r)} , -k , \frac{1}{r} , 0 \right)\nonumber\\
\hat{E}^{\mu'}_3 &=& \left( 0 , 0 , 0 , \frac{v_z}{c} \right) \label{t}
\end{eqnarray}
where $ k $ is a constant. The tangent vector to the geodesic, $ u^{\mu'} $ or $ \hat{E}^{\mu'}_+ $ takes the form
$ \hat{E}^{\mu'}_+ = (\dot{t},\dot{r},0,0) $, for a general affine parameter $ \lambda $. In the present case, we have set $ \dot{r} = 1 $, i.e., 
we have explicitly chosen $ r $ as the affine parameter along the geodesic $ \mathcal{N} $. This choice simplifies the calculation 
as well as it satisfies the required geodesic equation condition.

Now the above choice of tetrad has to satisfy the required conditions of Eq.(\ref{w}). Let us 
rewrite the first condition of Eq.(\ref{w}) and analyze it using the tetrad defined above. From the condition 
$\hat{E}_A \cdot \hat{E}_B = \eta_{AB}$, we obtain 
\begin{eqnarray}
&& \hat{E}_{-} \cdot \hat{E}_{-} = \eta_{--} = 0 ~~~~ ({\rm for} ~~ A=B=-) \nonumber \\
&& \Rightarrow ~  g_{\mu' \nu'} \hat{E}^{\mu'}_{-} \hat{E}^{\nu'}_{-} = 0 \Rightarrow ~ k(k-1) = 0 
\end{eqnarray}
So only two values of the constant $ k $ satisfy the required conditions for $ \hat{E}_{-} $, i.e $k$ takes two values,
$0$ and $1$. All other components of the tetrad 
automatically satisfy the required conditions of (\ref{w}). Therefore, we have two different set of tetrads with $ k = 0 $ and $ k = 1 $ 
which can be chosen as the basis of null Fermi frame.

Having obtained the Fermi tetrad basis, we can readily find out the components of the Riemann curvature tensor in null Fermi
coordinates using Eqs. (\ref{y}) and (\ref{t}). Then we use Eqs.(\ref{ai}) and (\ref{Ki}) 
to calculate the components of the effective magnetic field 
due to curvature coupling and the corresponding expressions are given by
\begin{eqnarray}
B^{+} &=& B^{-} = B^2 = 0 \nonumber\\
B^3 &=& \frac{k r (4 h-k y) v'(r)^2+v(r) \left(k r (4 h-k y) v''(r)+(k (-4 h+2 k y-y)+y)
   v'(r)\right)}{24 c^2 r}\nonumber\\
&~&
\label{u}
\end{eqnarray}
where $ h $, $ y $, $ z $ represent observer's coordinates or Fermi coordinates.

The above expression of $ B^3 $ has been evaluated for a general $ v(r) $. But in case of the draining bathtub type geometry 
discussed earlier, the specific form of $ v(r) $ happens to be $ v(r) = -\frac{cr_h}{r} $. So if we put this form of $ v(r) $ in Eq.(\ref{u}), 
we obtain the following expression of $ B^3 $ :
\begin{equation}
B^3 = \frac{r_h^2 \left[ 16 h k + \left(-1 + k -5 k^2 \right) y \right]}{24 r^4} \label{v}
\end{equation}
The expressions of $ B^3 $ for $ k=0 $ and $ k=1 $ are given by
\begin{equation}
B^3 = -\frac{r_h^2 y}{24 r^4} ~~(k=0) ~ ,~B^3 = \frac{r_h^2 (16 h-5 y)}{24 r^4} ~~(k=1)
\label{B3onetwo}
\end{equation}
We need to analyze this result in more detail. Eq.(\ref{v}) tells us that the effective magnetic field component ($ b^3 $) 
diverges at $ r = 0 $. Since $ r = r_h $ represents the position of event horizon of the analog black hole and we are particularly 
interested in the phenomena occurring outside $ r_h $, the effective magnetic field is always finite in this region. From Eq.(\ref{v}) 
we see that the effective magnetic field falls off as $ r^{-4} $ as a function of the radial distance. 

\section{Massless fermionic quasiparticles in circular null geodesic}

We will now compute the curvature coupling of fermionic quasiparticles in circular null geodesics. This is of course a special
case, as we discuss. For such geodesics, it can be checked from Eq.(\ref{c}) that the only allowed value of the radial
coordinate is $r = \sqrt{2}r_h$, for $v(r) = -cr_h/r$ as in the previous subsection. 
This is the analog of the photon sphere in GR \cite{Virbhadra}, and a null observer in a circular geodesic 
is uniquely located at this value of $r$. This is to be kept in mind in the analysis that follows. 

Similar to the radial case, first we need to set up the pseudo-orthonormal Fermi frame for a circular null geodesic 
$ \mathcal{G} $. Then we find out components of Riemann tensor in Fermi coordinates and use it to calculate the effective 
magnetic field. By `circular null geodesic' we mean the family of null geodesics for which $ r $ is constant, i.e., 
$ \dot{r} = \ddot{r} = 0 $. The corresponding Fermi frame for a circular null geodesic $ \mathcal{G} $ are found to be
\begin{eqnarray}
\hat{E}^{\mu'}_+ &=& \left( \frac{1}{\sqrt{c^2-v(r)^2}} , 0 , \frac{1}{r} , 0 \right) \nonumber\\
\hat{E}^{\mu'}_- &=& \left(\frac{r \phi ^2 v(r) v'(r)}{2 c^2 \sqrt{c^2-v(r)^2}}+\frac{\phi  v(r) }{c^2}
-\frac{1}{2  \sqrt{c^2-v(r)^2}}, \right. \nonumber\\ &~&\left. -\frac{ r\phi  v(r) v'(r)}{c^2}, 
   \frac{ \phi ^2 v(r) v'(r)}{2 c^2 }+\frac{1}{2 r},0\right)\nonumber\\
\hat{E}^{\mu'}_2 &=& \left( -\frac{\phi }{c} + \frac{v(r)}{c \sqrt{c^2-v(r)^2}}, \frac{\sqrt{c^2-v(r)^2}}{c} , -\frac{\phi  \sqrt{c^2-v(r)^2}}{c r} , 0 \right)\nonumber\\
\hat{E}^{\mu'}_3 &=& \left( 0 , 0 , 0 , \frac{v_z}{c} \right) \label{al}
\end{eqnarray}
with the condition $rv(r)v'(r) + c^2 - v(r)^2 = 0$. 
It has to be remembered that the above expressions in Eq.(\ref{al}) have to be 
evaluated at $r = \sqrt{2}r_h$ and we have denoted 
$ v'(r) = \frac{\partial v(r)}{\partial r} $.

It is to be noted that the tetrad components depend explicitly on $\phi$. This might seem surprising, given that
the analog metric that we start with is spherically symmetric, but is due to the fact that the tetrad must satisfy 
the pseudo-orthonormal and parallel transport conditions given in Eq.(\ref{w}),
along the null geodesic. As the first two vectors of the tetrad basis $\hat{E}^+$ and $\hat{E}^{-}$ are 
null, for circular geodesic they demand its components to depend explicitly on the affine parameter which in this case
is chosen to be the arc-length of the circular geodesic $\sqrt{2} r_h \phi$. Similar dependence can also be seen for the 
timelike circular geodesic where $\phi$ dependence comes into the phases of harmonic functions \cite{13}
but not explicitly, the reason being that the tetrad basis is made of the timelike tangent vector and three spacelike vectors. 

The corresponding expressions of the components of effective magnetic field in this case are
\begin{eqnarray}	
B^{+} &=& B^{-} = B^2 = 0 \nonumber\\
B^3 &=& \frac{8 \sqrt{2}  h \phi  \left(\phi ^2+4\right)- \left(\phi ^4-96\right)y}{384  r_h^2} 
\label{an} 
\end{eqnarray}

\section{Numerical Estimates}

It now remains to provide numerical estimates of the $B^A$ that we have evaluated. In order to do this, we will take recourse to
various approximations that we now discuss. We note that the dimension of $B^A$ for both radial and circular geodesics
is an inverse length (contrary to usual magnetic fields that come in dimensions of $1/L^2$). In order to convert $B^A$ into
a quantity having dimensions of a magnetic field (Gauss or Tesla), we will need to divide it by the Bohr magneton, expressed
in GeV per Tesla (or GeV per Gauss) \cite{Bluhm}. Doing this, it can be checked that a magnetic field of  $10^{-12}$ Gauss translates to 
a value of $B^A \sim 10^{-29}~{\rm GeV}$. This is the limit of measurability of the magntic field, as of now. In the analysis
that follows, we will use energy units only, for convenience. 

Let us first consider the case of radial geodesics and consider the case $k=0$, i.e 
the first expression given in Eq.(\ref{B3onetwo}). Since this has dimension $L^{-1}$, we convert this to energy units by 
multiplying with $\hbar c_L$, where $c_L$ is the speed of light $(=3 \times 10^8~{\rm m/s})$. Thus we have 
$B^3 (k=0) = - r_h^2 y\hbar c_L/(24 r^4)$. In order to get an estimate for the coordinate $y$, we use the uncertainty
relation. Remembering that the quasiparticles are dressed Helium-3 atoms of mass $m^*$, moving with speed $c = 3 {\rm~cm/s}$, 
we have $y \sim \hbar/p = \hbar / (m^*c)$. Plugging this in, we have in electron-volts, 
\begin{equation}
|B^3 (k=0)| = \frac{1}{24}\frac{r_h^2\hbar^2c_L}{r^4 m^* c\times q_e}~~{\rm eV}
\label{Beq}
\end{equation}
where $q_e$ is the electron charge. We now use the fact that $m^*$ is 3 times the mass of Helium-3 atoms, which is
given by $3.016$ atomic mass units. A numerical estimate of $|B^3 (k=0)|$ is obtained by putting in this value of $m^*$. 

Now we note that the theory of low energy quasiparticles is valid for 
\begin{equation}
E \ll \frac{\Delta_v^2}{p_Fv_F}~, {\rm i.e}~~ E \ll 10^{-10}~{\rm eV}
\label{Eeq}
\end{equation}
Since $B^A$ appears in the interaction energy term in the Lagrangian of Eq.(\ref{Lag1}), as an estimate we 
equate $B^3 (k=0) \sim 10^{-10}$, to obtain $r = 0.067\sqrt{r_h}$ (in metres). For $E \ll 10^{-10} {\rm eV}$, we therefore
require $r \gg 0.067\sqrt{r_h}$. A typical value of $r_h \sim 1~\mu {\rm m}$ will thus ensure that our results are valid 
for $r \gg 0.0067~{\rm cm}$. Hence on a radial geodesic, at say $r \sim 1~{\rm cm}$, our results should be robust, and at
this radial distance, we have $B^3 (k=0) \sim 10^{-28}~{\rm GeV}$. Smaller values of the radial distance may push up this
value to $\sim 10^{-27} - 10^{-26}~{\rm GeV}$, while respecting the energy condition. 
This discussion was for $k=0$. For $k=1$, the analysis of the second expression of Eq.(\ref{B3onetwo}) is qualitatively similar, and 
yields similar numerical estimates. 

We will now turn to circular geodesics. An analysis similar to the one outlined above shows that in this case, setting
$y=0$ in Eq.(\ref{an}) implies that $B^3 = (1/r_h^2)(5.6 \times 10^{-15}\phi + 1.4 \times 10^{-15} \phi^3)~{\rm eV}$. 
If we now set a typical value of $\phi = \pi$, then in order to satisfy $E \ll 10^{-10} {\rm eV}$,
we require $r_h \gg 2.4~{\rm cm}$. As an estimate, if we set $r_h = 10~{\rm cm}$, we obtain $B^3 \sim 10^{-21}~{\rm GeV}$.

Our numerical analysis above establishes the fact that the effective magnetic field that is seen by an inner observer
in superfluid $\text{He}^3$-A are withing bounds reachable by present experiments, i.e these are not vanishingly small.
Hence, such an observer should measure effects that are known in quantum mechanics regarding the interaction of spins
with such magnetic fields. In this case, however, the effective magnetic field is non-uniform. For radial null geodesics, this 
falls off as the fourth power of the radial distance, while for circular null geodesics, it is explicitly dependent on the angular
variable. From Eq.(\ref{B3onetwo}), using the condition $r \gg 0.067\sqrt{r_h}$, it is seen that for relatively large values of
$r$ (compared to $r_h$), $B^3$ varies slowly. Hence, if we approximate $B^3$ by a uniform (average) value, for such large
$r$, one should expect the inner observer to see oscillations 
between a spin up and a spin down state of the massless fermionic quasiparticles when the system evolves from
a general spin state. Similar analysis holds for null circular
geodesics, for small values of the angular coordinate. The external observer (moving at $3$ cm per
sec) along a radial coordinate or moving at a fixed radius, however, perceives these quasiparticles as dressed Helium-3 atoms. 
It would therefore seem that such an observer is likely to see the spin of the quasiparticles to also oscillate between an up-spin
state and a down-spin state. In fact, this can be quantified further. 

The time difference between two events for the external observer $\Delta t$ is related to that for the inner observer $\Delta\tau$
by \cite{VolovikBookHelium} $\Delta t = \Delta\tau/\sqrt{1-r_h^2/r^2}$. Hence, the characteristic frequency of oscillation
(assuming a uniform magnetic field) for the external observer is dilated by a factor of $\sqrt{1-r_h^2/r^2}$, as compared to
the inner observer. For null radial geodesics, for small values of $r_h/r$, $\Delta t \sim \Delta\tau$. This is relevant for us, as
we have already seen that the energy condition demands that here, $r \gg 0.067\sqrt{r_h}$, and that the effective magnetic
field can be approximated to a constant for small values of $r_h/r$.
For circular geodesics, since $r = \sqrt{2}r_h$, the dilation factor is $\sqrt{2}$. 
Possible experimental signatures of this might be an interesting issue to pursue, modulo a
limitation of our analysis that we point out in the next section. 

\section{Discussions, conclusions and limitations}

In this paper, we started with a $ \text{He}^3 $-superfluid system where the vacuum excitations are Bogoliubov fermions, 
which are dressed $ \text{He}^3 $ atoms and see an effective curved space-time with moving superfluid vacuum in the background. 
We first established the Fermi coordinates along a null radial as well as a null circular geodesics, and calculated the components of 
the Riemann curvature tensor in these coordinates. 
Having obtained the curvature tensor, we determined an effective magnetic field due to curvature coupling. The whole analysis
was done in the analog black hole draining bath-tub geometry of fig.(\ref{fg1}) discussed by Volovik in \cite{5}. 

As we have mentioned in the beginning of this paper, spinning particles do not follow exactly geodesic trajectories, but 
the difference of the latter from their actual paths is small. We can however envisage an internal null observer (feeling the 
analog metric), on such a geodesic trajectory, who makes a measurement on the fermionic system. This internal null observer
however moves with a finite speed ($\sim 3~{\rm cm/s}$) and sees the non-trivial effect of curvature coupling to the 
fermionic quasiparticles. It is well known (see, e.g \cite{Sakurai}) that spin half fermions that interact with a constant external magnetic
field oscillates between the up and down states as it evolves from a general spin state. 
The frequency of this oscillation is proportional to the Bohr magneton. In 
this case, we obtain an effective interaction term that is similar to the former, and if we approximate the magnetic field to 
a uniform value assuming that it varies slowly, similar effects of oscillation between the
up and down spin states should result from a standard quantum mechanical analysis. 
It is not difficult to identify an external observer who moves in a 
circular or radial trajectory in the given geometry, and coincides with the internal observer. Our analysis
would imply that this external observer should see the same effects on the spin state of the dressed Helium-3 atom that
she perceives, and this might be a measurable phenomenon in futuristic experiments. 

Before ending this paper, we should point out that the analysis that we have presented here is limited by the fact that
it is applicable only to two special class of geodesics, i.e radial or circular. A generic geodesic path may be neither of these. 
However, this last case is difficult to analyse analytically, and we leave a study of such a situation for a future publication. 

\begin{center}
{\bf Acknowledgements}
\end{center}
\noindent
It is a pleasure to thank K. Bhattacharya and V. Subhrahmanyam for useful discussion. 

\bibliographystyle{utphys}
\bibliography{References}

\providecommand{\href}[2]{#2}\begingroup\raggedright\begin{thebibliography}{10}

\bibitem{Weinberg}
S.~Weinberg, {\em {Gravitation and Cosmology}}.
\newblock John Wiley and Sons, New York,
1972.
\newblock

\bibitem{Wald}
R.~M. Wald, {\em General relativity}.
\newblock University of Chicago press, 2010.

\bibitem{Hartle}
J.~B. Hartle, {\em Gravity: An introduction to Einstein’s general
  relativity}.
\newblock AAPT, 2003.

\bibitem{7}
E.~Poisson, {\em A relativist's Toolkit, The Mathematics of Black-Hole
  Mechanics.}
\newblock Cambridge University Press, 2004.

\bibitem{GPB}
C.~W.~F. Everitt {\em et~al.}, ``{Gravity Probe B: Final Results of a Space
  Experiment to Test General Relativity},''
  \href{http://dx.doi.org/10.1103/PhysRevLett.106.221101}{{\em Phys. Rev.
  Lett.} {\bfseries 106} (2011) 221101},
\href{http://arxiv.org/abs/1105.3456}{{\ttfamily arXiv:1105.3456 [gr-qc]}}.

\bibitem{1}
W.~G. Unruh, ``Experimental black-hole evaporation?,''
  \href{http://dx.doi.org/10.1103/PhysRevLett.46.1351}{{\em Phys. Rev. Lett.}
  {\bfseries 46} (May, 1981) 1351--1353}.

\bibitem{2}
M.~Visser, ``{Acoustic black holes},'' in {\em {Advanced School on Cosmology
  and Particle Physics (ASCPP 98) Peniscola, Spain, June 22-28, 1998}}.
\newblock
\href{http://arxiv.org/abs/gr-qc/9901047}{{\ttfamily arXiv:gr-qc/9901047
  [gr-qc]}}.
\newblock

\bibitem{3}
T.~A. Jacobson and G.~E. Volovik, ``{Event horizons and ergoregions in He-3},''
\href{http://dx.doi.org/10.1103/PhysRevD.58.064021}{{\em Phys. Rev. D}
  {\bfseries 58} (1998) 064021}.

\bibitem{analog1}
C.~Barcelo, S.~Liberati, and M.~Visser, ``{Analogue gravity},''
  \href{http://dx.doi.org/10.12942/lrr-2005-12}{{\em Living Rev. Rel.}
  {\bfseries 8} (2005) 12},
\href{http://arxiv.org/abs/gr-qc/0505065}{{\ttfamily arXiv:gr-qc/0505065
  [gr-qc]}}.

\bibitem{analog2}
S.~Weinfurtner, S.~Liberati, and M.~Visser, ``{Analogue model for quantum
  gravity phenomenology},''
  \href{http://dx.doi.org/10.1088/0305-4470/39/21/S83}{{\em J. Phys.}
  {\bfseries A39} (2006) 6807--6814},
\href{http://arxiv.org/abs/gr-qc/0511105}{{\ttfamily arXiv:gr-qc/0511105
  [gr-qc]}}.

\bibitem{analog3}
D.~V. Fursaev, ``{Entanglement entropy in critical phenomena and analogue
  models of quantum gravity},''
  \href{http://dx.doi.org/10.1103/PhysRevD.73.124025}{{\em Phys. Rev.}
  {\bfseries D73} (2006) 124025},
\href{http://arxiv.org/abs/hep-th/0602134}{{\ttfamily arXiv:hep-th/0602134
  [hep-th]}}.

\bibitem{analog4}
T.~K. Das, N.~Bilic, and S.~Dasgupta, ``{Black-Hole Accretion Disc as an
  Analogue Gravity Model},''
  \href{http://dx.doi.org/10.1088/1475-7516/2007/06/009}{{\em JCAP} {\bfseries
  0706} (2007) 009},
\href{http://arxiv.org/abs/astro-ph/0604477}{{\ttfamily arXiv:astro-ph/0604477
  [astro-ph]}}.

\bibitem{analog5}
G.~Krein, G.~Menezes, and N.~F. Svaiter, ``{Analog model for quantum gravity
  effects: Phonons in random fluids},''
  \href{http://dx.doi.org/10.1103/PhysRevLett.105.131301}{{\em Phys. Rev.
  Lett.} {\bfseries 105} (2010) 131301},
\href{http://arxiv.org/abs/1006.3350}{{\ttfamily arXiv:1006.3350 [hep-th]}}.

\bibitem{analog6}
S.~Hossenfelder, ``{Analog Systems for Gravity Duals},''
  \href{http://dx.doi.org/10.1103/PhysRevD.91.124064}{{\em Phys. Rev.}
  {\bfseries D91} no.~12, (2015) 124064},
\href{http://arxiv.org/abs/1412.4220}{{\ttfamily arXiv:1412.4220 [gr-qc]}}.

\bibitem{analog7}
B.~Cropp, S.~Liberati, and R.~Turcati, ``{Vorticity in analog gravity},''
  \href{http://dx.doi.org/10.1088/0264-9381/33/12/125009}{{\em Class. Quant.
  Grav.} {\bfseries 33} no.~12, (2016) 125009},
\href{http://arxiv.org/abs/1512.08198}{{\ttfamily arXiv:1512.08198 [gr-qc]}}.

\bibitem{ParthaDa}
C.~Chakraborty, O.~Ganguly, and P.~Majumdar, ``{Intertial Frame Dragging in an
  Acoustic Analogue spacetime},''
\href{http://arxiv.org/abs/1510.01436}{{\ttfamily arXiv:1510.01436 [gr-qc]}}.

\bibitem{11}
L.~Parker, ``One-electron atom as a probe of spacetime curvature,''
  \href{http://dx.doi.org/10.1103/PhysRevD.22.1922}{{\em Phys. Rev. D}
  {\bfseries 22} (Oct, 1980) 1922--1934}.

\bibitem{Parker2}
L.~Parker, ``{One-Electron Atom in Curved Space-Time},''
\href{http://dx.doi.org/10.1103/PhysRevLett.44.1559}{{\em Phys. Rev. Lett.}
  {\bfseries 44} no.~23, (1980) 1559}.

\bibitem{12}
L.~Parker and L.~O. Pimentel, ``Gravitational perturbation of the hydrogen
  spectrum,'' \href{http://dx.doi.org/10.1103/PhysRevD.25.3180}{{\em Phys. Rev.
  D} {\bfseries 25} (Jun, 1982) 3180--3190}.

\bibitem{13}
A.~Dey, A.~Samanta, and T.~Sarkar, ``Fermi normal coordinates and fermion
  curvature couplings in general relativity,''
  \href{http://dx.doi.org/10.1103/PhysRevD.89.104008}{{\em Phys. Rev. D}
  {\bfseries 89} (May, 2014) 104008}.

\bibitem{8}
S.~Mohanty, B.~Mukhopadhyay, and A.~R. Prasanna, ``Experimental tests of
  curvature couplings of fermions in general relativity,''
  \href{http://dx.doi.org/10.1103/PhysRevD.65.122001}{{\em Phys. Rev. D}
  {\bfseries 65} (May, 2002) 122001}.

\bibitem{9}
U.~Debnath, B.~Mukhopadhyay, and N.~Dadhich, ``Spacetime curvature coupling of
  spinors in early universe: Neutrino asymmetry and a possible source of
  baryogenesis,'' \href{http://dx.doi.org/10.1142/S0217732306019542}{{\em
  Modern Physics Letters A} {\bfseries 21} no.~05, (2006) 399--408}.

\bibitem{6}
F.~K. Manasse and C.~W. Misner, ``Fermi normal coordinates and some basic
  concepts in differential geometry,''
  \href{http://dx.doi.org/10.1063/1.1724316}{{\em Journal of Math. Phys.}
  {\bfseries 4} no.~6, (1963) 735--745}.

\bibitem{YeeBander}
K.~Yee and M.~Bander, ``{Equations of motion for spinning particles in external
  electromagnetic and gravitational fields},''
  \href{http://dx.doi.org/10.1103/PhysRevD.48.2797}{{\em Phys. Rev.} {\bfseries
  D48} (1993) 2797--2799},
\href{http://arxiv.org/abs/hep-th/9302117}{{\ttfamily arXiv:hep-th/9302117
  [hep-th]}}.

\bibitem{VolovikBookHelium}
G.~Volovik, {\em The universe in a helium droplet, volume 117 of International
  series of monographs on physics}.
\newblock Clarendon Press, Oxford, 2003.

\bibitem{5}
G.~Volovik, ``Simulation of a panlevé-gullstrand black hole in a thin [3]he-a
  film,''
  \href{http://dx.doi.org/http://cds.cern.ch/record/377679/files/9901077.pdf}{{\em
  JETP LETTERS} {\bfseries 69} no.~9, (1999) 705--713}.

\bibitem{4}
M.~Visser, ``{Acoustic black holes: Horizons, ergospheres, and Hawking
  radiation},'' \href{http://dx.doi.org/10.1088/0264-9381/15/6/024}{{\em Class.
  Quant. Grav.} {\bfseries 15} (1998) 1767--1791},
\href{http://arxiv.org/abs/gr-qc/9712010}{{\ttfamily arXiv:gr-qc/9712010
  [gr-qc]}}.

\bibitem{10}
V.~Kostelecky and C.~Lane {\em J. Math. Phys. (N.Y.)} {\bfseries 40} (1999)
  6245.

\bibitem{16}
M.~Blau, D.~Frank, and S.~Weiss, ``{Fermi coordinates and Penrose limits},''
  \href{http://dx.doi.org/10.1088/0264-9381/23/11/020}{{\em Class. Quant.
  Grav.} {\bfseries 23} (2006) 3993--4010},
\href{http://arxiv.org/abs/hep-th/0603109}{{\ttfamily arXiv:hep-th/0603109
  [hep-th]}}.

\bibitem{VolovikWeyl}
G.~Volovik, ``{Analog of gravity in superfluid ${}^3\text{He}$-A},'' {\em JETP
  Lett} {\bfseries 44} no.~8, (1986) .

\bibitem{Virbhadra}
C.-M. Claudel, K.~S. Virbhadra, and G.~F.~R. Ellis, ``{The Geometry of photon
  surfaces},'' \href{http://dx.doi.org/10.1063/1.1308507}{{\em J. Math. Phys.}
  {\bfseries 42} (2001) 818--838},
\href{http://arxiv.org/abs/gr-qc/0005050}{{\ttfamily arXiv:gr-qc/0005050
  [gr-qc]}}.

\bibitem{Bluhm}
R.~Bluhm and V.~A. Kostelecky, ``{Lorentz and CPT tests with spin polarized
  solids},'' \href{http://dx.doi.org/10.1103/PhysRevLett.84.1381}{{\em Phys.
  Rev. Lett.} {\bfseries 84} (2000) 1381--1384},
\href{http://arxiv.org/abs/hep-ph/9912542}{{\ttfamily arXiv:hep-ph/9912542
  [hep-ph]}}.

\bibitem{Sakurai}
J.~J. Sakurai and E.~D. Commins, ``Modern quantum mechanics, revised edition,''
  1995.

\end{thebibliography}\endgroup
\end{document}